\pdfoutput=1  
\documentclass[]{article}  
\usepackage{url,float}
\usepackage{graphicx}
\usepackage{amsmath}
\usepackage{amsfonts}
\usepackage{amssymb}
\usepackage{latexsym}

\newcommand{\hide}[1]{}

\newcommand{\ABox}{
\raisebox{3pt}{\framebox[6pt]{\rule{6pt}{0pt}}}
}
\newenvironment{proof}{{\bf Proof:}}{\hfill\ABox}

\newtheorem{theorem}{{\bf Theorem}}

\newtheorem{lemma}{Lemma}

\newcommand{\lemlab}[1]{\label{lemma:#1}}
\newcommand{\thmlab}[1]{\label{thm:#1}}

\newcommand{\figlab}[1]{\label{fig:#1}}
\newcommand{\seclab}[1]{\label{sec:#1}}
\newcommand{\figref}[1]{\ref{fig:#1}}

{\makeatletter
 \gdef\xxxmark{%
   \expandafter\ifx\csname @mpargs\endcsname\relax 
     \expandafter\ifx\csname @captype\endcsname\relax 
       \marginpar{xxx}
     \else
       xxx 
     \fi
   \else
     xxx 
   \fi}
 \gdef\xxx{\@ifnextchar[\xxx@lab\xxx@nolab}
 \long\gdef\xxx@lab[#1]#2{{\bf [\xxxmark #2 ---{\sc #1}]}}
 \long\gdef\xxx@nolab#1{{\bf [\xxxmark #1]}}
 \gdef\turnoffxxx{\long\gdef\xxx@lab[##1]##2{}\long\gdef\xxx@nolab##1{}}%
}

\def\R{{\mathbb{R}}}

\def\C{{\mathcal{C}}}

\def\R{{\mathbb{R}}}

\title{%
On Folding a Polygon to a Polyhedron
} 

\author{%
Joseph O'Rourke%
    \thanks{Department of Computer Science, Smith College, Northampton, MA
      01063, USA.
      \protect\url{orourke@cs.smith.edu}.}
}

\begin{document}
\maketitle

\begin{abstract}
We show that the open problem presented in 
\emph{Geometric Folding Algorithms: Linkages, Origami, Polyhedra}~\cite{do-gfalop-07}
is solved by a theorem of Burago and Zalgaller~\cite{bz-ipli2-96}
from more than a decade earlier.
\end{abstract}

\section{Introduction}
\seclab{Introduction}

In~\cite[p.~384]{do-gfalop-07}
we formulated this open problem:

\begin{center}
\vspace{1mm}
\fbox{
\begin{minipage}[h]{0.95\linewidth}
\noindent
\textsc{Open Problem~25.1:} 
Folding Polygons to (Nonconvex) Polyhedra \\

Does every simple polygon fold (by perimeter gluing) to some simple polyhedron?
In particular, does some perimeter halving always lead to a polyhedron?

\end{minipage}
}
\vspace{1mm}
\end{center}

This note explains how this problem is solved (positively) by a
theorem in~\cite{bz-ipli2-96}.
There is nothing original in this note.
It is merely a description of earlier work,
interpreting it in the context of the folding problem.

The theorem of Burago and Zalgaller~\cite[p.~370]{bz-ipli2-96} that solves the
problem is this:

\begin{theorem}[Burago-Zalgaller~1.7]
``Every polyhedron $M$ admits an isometric piecewise-linear $\C^0$
immersion
into $\R^3$.  If $M$ is orientable or has a nonempty boundary,
then $M$ admits an  isometric piecewise-linear $\C^0$
embedding
into $\R^3$.''
\end{theorem}

Interpreting this theorem (henceforth, the BZ theorem) and connecting it to the folding problem
requires an analysis of its
technical terms.

\section{Definition of Terms}
\seclab{Definitions}

First we explain the folding problem from~\cite{do-gfalop-07}.
Folding a polygon by perimeter gluing means identifying points of
the boundary of the polygon with other points so that the resulting
manifold is topologically a sphere, that is, closed, orientable, and
with genus zero.
Isolated points might have no match, which occurs where the
boundary is ``zipped'' in the neighborhood of such a point.
This is perhaps clearest for a perimeter halving.
Let $x$ and $y$ be points on the boundary of the polygon that
partition
the perimeter into two equal-length halves.  Then  
\emph{perimeter halving} identifies the two halves of the boundary,
zipping in the neighborhoods of $x$ and $y$.
(See ahead to Figure~\figref{socgPoly}.)
Our question was whether every perimeter gluing, and more
specifically,
every perimeter halving, produces (is realized by) a simple (non-self-intersecting)
polyhedron.

This question is an analog of Alexandrov's 1941 theorem,
described in~\cite[Sec.~23.3]{do-gfalop-07},
\cite[Sec.~37]{p-ldpg-10}, and Alexandrov's book~\cite{a-cp-05}.
Alexandrov's theorem adds one more condition---that the gluing creates no more than $2 \pi$ surface 
angle surrounding any point
of the resulting manifold---and 
concludes that the result is a unique \emph{convex} polyhedron,
where ``polyhedron'' is interpreted to include a flat doubly covered
convex polygon.
The focus of our problem was to ask for an extension without the $2
\pi$ convexity condition.
The restriction to folding a single polygon is not essential to the
problem.
Alexandrov considered a more general gluing of a collection of
polygons,
which has been translated as a gluing of a
``development''~\cite[p.~50ff]{a-cp-05},
terminology not always followed by later authors.

\begin{sloppypar}
We will draw several definitions from~\cite{p-ips-06}
(many also incorporated into~\cite{p-ldpg-10}),
for they seem the clearest on this topic.
In particular, the notion of a ``gluing''  is captured
in this definition~\cite[Sec.~1.3]{p-ips-06}:
\end{sloppypar}

\begin{quotation}
\noindent
``Let $S$ be an \emph{abstract 2-dimensional polyhedral surface}
defined as a collection of
triangles $T_1,\ldots,T_m$ with combinatorial gluing rules. Here each triangle $T_i$ is given
by its edge lengths, and whenever two edges are glued, they have equal
length.
We
always assume that $S$ is a connected simplicial complex, and that it is closed (has no
boundary) and orientable, i.e. homeomorphic to a sphere with $g \ge 0$
handles.''
\end{quotation}

\noindent
Restricting to $g=0$ yields exactly the gluings that are our main
focus.
Our polygon can be partitioned into triangles so that the perimeter
gluing is captured by edge-to-edge gluings of the triangles.
To simplify the language, we will call Pak's ``abstract 2-dimensional
polyhedral surface''
a \emph{gluing of polygons}.
Next we concentrate on the phrase
``every polyhedron $M$'' in the BZ theorem,
and show that it refers to gluings of
polygons.
From~\cite [p.~369]{bz-ipli2-96}:
\begin{quotation}
\noindent
``By a two-dimensional manifold with polyhedral metric (in brief, a
\emph{polyhedron}),
we mean a metric space endowed with the structure of a connected
compact
two-dimensional manifold (possibly with boundary) every point $x$ of
which
has a neighborhood isometric to the vertex of a cone. ...
The metric is locally flat everywhere except for a finite collection
of points; these points are the `true' vertices.''
\end{quotation}
Although their naming this concept a ``polyhedron'' is nonstandard,
it is clear this coincides with the notion of a gluing of polygons.
The ``true vertices'' $V$ of such a gluing are the points whose total
angle
differs from $2 \pi$, at which the discrete curvature is concentrated.

\begin{sloppypar}
For definitions of ``immersion'' and ``embedding,''
we again turn to Pak~\cite{p-ips-06}:
\end{sloppypar}

\begin{quotation}
\noindent
``A (3-dimensional) \emph{realization} of $S$ is defined as a map $f: V
\mapsto \R^3$  such that the
Euclidean distance $||v_1,v_2||$ between vertices is equal to the edge length $|v_1,v_2|$ of any
triangle $T_i$ which contains $v_1$ and $v_2$.

An \emph{immersion} is a realization where no two triangles have a
2-dimensional intersection. 
For example, a doubly covered triangle is a realization in $\R^3$ of a surface
homeomorphic to a sphere, but not an immersion.
An \emph{embedding} is a realization where two triangles intersect only by an edge or by
a vertex they share. We always consider surfaces $S$ up to isometry, so we speak of
isometric immersions and isometric embeddings.''
\end{quotation}

Our interest in the folding problem is focused on embeddings (with the
exception of doubly covered polygons):
because we imagine performing the folding with paper, we do not
want the paper to penetrate itself.
So we henceforth concentrate on the second half of the BZ theorem,
which describes embeddings.

In our context, that the embedding is $\C^0$ simply means that it is
continuous,
that is, without tearing.
I believe they include this qualification primarily to distinguish
their result from the famous $\C^1$-smooth embedding theorem
of Nash~\cite{n-c1ii-54} (improved by Kuiper~\cite{k-oc1ii-55}).
This Nash-Kuiper result cannot be improved to $\C^2$.
Histories of immersion and embedding theorems can be found 
in~\cite{s-gaitt-05} and~\cite{s-imivf-10}.

The key phrase in the BZ theorem is that the embedding is
``isometric piecewise-linear.''
Isometric simply means distances within the surface are maintained.
Piecewise-linear means that each triangle $T_i$ from the gluing of
polygons is mapped to a finite collection of triangles in $\R^3$.
Continuous isometric deformations are often called
\emph{bendings}  in the literature (e.g.,
\cite[Sec.~38.1]{p-ldpg-10}).
The addition of ``piecewise-linear'' means that the bending is
accomplished
with flat pieces.
A nice phrase used in~\cite [Lem.~2.2]{bz-ipli2-96}
to describe the mapping of one triangle is that it becomes a
```pleated' surface'' in $\R^3$.
The triangle gets creased along a network of straight segments.
From the construction in that paper, it is clear that
only a finite number of pieces are used to embed each triangle
$T_i$.

So we may rephrase the embedding part of the BZ theorem that is our
focus 
as follows:
\begin{theorem}
Every gluing of polygons to form an oriented manifold
has an isometric embedding as a polyhedral surface
(composed of a finite number of flat triangles)
in $\R^3$.
\end{theorem}

Because perimeter gluing, and perimeter halving in particular,
constitutes a gluing of polygons, this theorem provides a positive
answer
to Open Problem~25.1.

I presented Figure~\figref{socgPoly} at the 20th \emph{Symposium on
  Computational Geometry} (the proceedings cover image) as a challenge to fold to a
polyhedron.
The figure marks out a particular perimeter-halving folding,
which, by  BZ's theorem, must fold to a simple polyhedron.
\begin{figure}[htbp]
\centering
\includegraphics[width=0.75\linewidth]{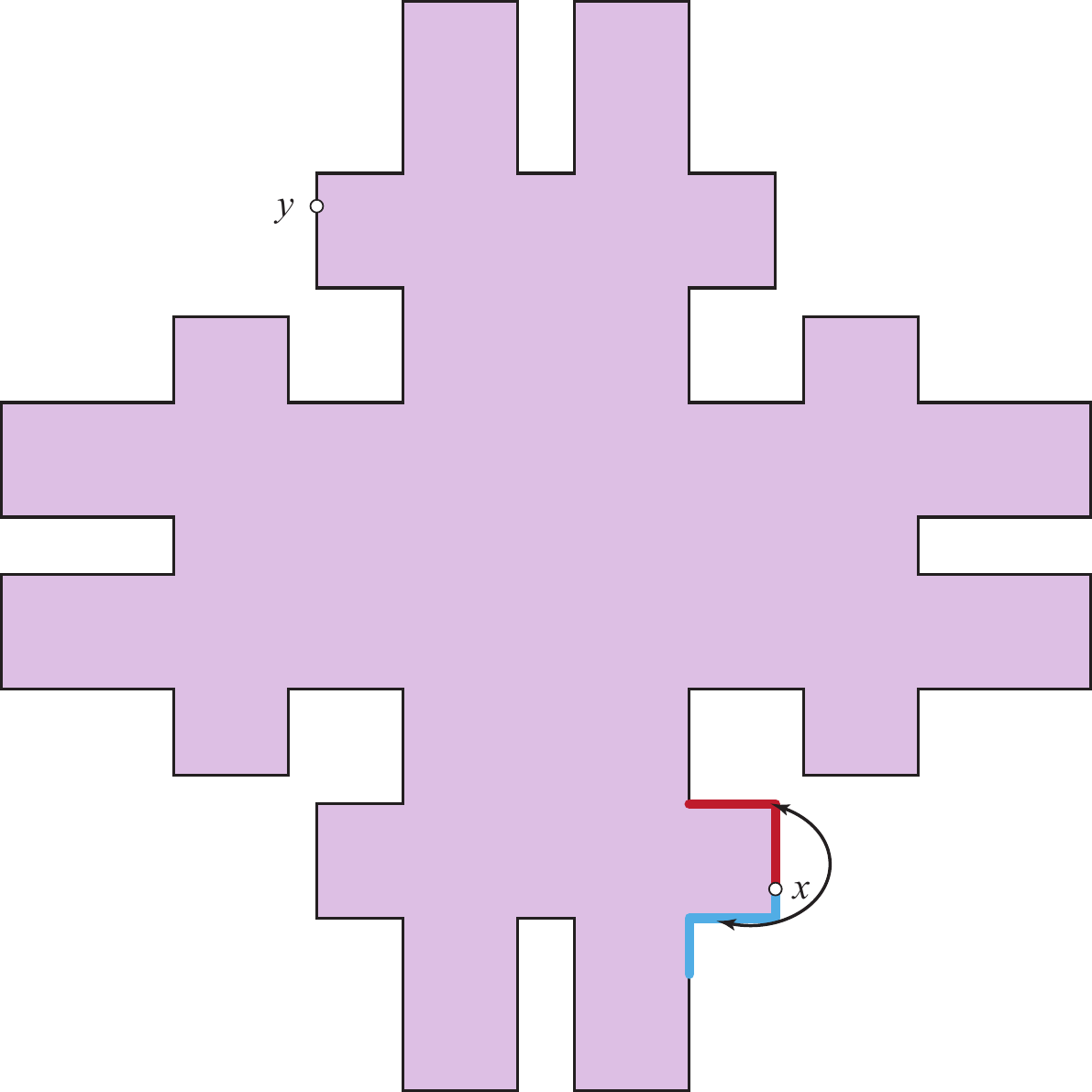}
\caption{Points $x$ and $y$ are perimeter-halving points.
The folding ``zips'' the perimeter closed from $x$ to $y$.
A short initial segment of the gluing
in the neighborhood of $x$ is indicated:
the red portion of the boundary glues to the symmetric blue portion.
This perimeter gluing continues from $x$ to $y$.}
\figlab{socgPoly}
\end{figure}

\section{Discussion}
\seclab{Discussion}

One of the motivations of the work in~\cite{bz-ipli2-96}
was to show that one can increase the volume of a convex polyhedron
by making it nonconvex, but otherwise remaining isometric to the
original.
This question is explored in great depth in~\cite{p-ips-06}
and~\cite{p-icwos-08}.
The constructions are piecewise-linear, that is, polyhedral.

Although I have been citing~\cite{bz-ipli2-96} as resolving the
folding
problem, in fact earlier (as yet untranslated) work of
Burago and
Zalgaller~\cite{bz-pen-60}
apparently already sufficed for the special case when
the manifold is homeomorphic to a sphere.
This is cited in support of Exercise~39.13(a)
in~\cite{p-ldpg-10}.

Although the problem is solved by the BZ theorem,
their proof is sufficiently complex that it is difficult to
see what their construction will produce for a specific example,
such as the folding in Figure~\figref{socgPoly}.
Saucan has a useful summary of the Burago-Zalgaller construction
in~\cite{s-imivf-10}.
The construction incorporates ideas from Nash's $\C^1$ embedding
construction.
In particular, Nash's ```spiralling' perturbations''~\cite[p.~383]{n-c1ii-54} of the surface results in a
polyhedral
surface that is ``strongly `corrugated','' to quote Saucan's apt
description.
Although the number of triangular facets of the final embedded
polyhedron
is finite, it does not seem straightforward to provide an explicit upper bound
on the number of facets.
The work of Bern and Hayes in~\cite{bh-oeplt-08} achieves a flat
embedding in $\R^2$ with just $O(n)$ facets, where $n$ is the
number of triangles of the polygon gluing.  
Perhaps their work will lead
to a more efficient construction in $\R^3$.



\paragraph{Acknowledgments.}
I thank 
Igor Pak for connecting me to the
relevant literature, and patiently explaining key definitions.
Any remaining misinterpretations are solely my own.

\bibliographystyle{alpha}
\bibliography{/Users/orourke/bib/geom/geom}
\end{document}